\documentclass[%
 aip,
 amsmath,amssymb,
]{revtex4-1}

\usepackage{graphicx}
\usepackage{dcolumn}
\usepackage{bm}
\usepackage[utf8]{inputenc}
\usepackage[T1]{fontenc}
\usepackage{mathptmx}
\usepackage{xcolor}
\usepackage{anyfontsize}
\usepackage{t1enc}
\usepackage{epstopdf}
\usepackage{makecell}
\newcommand{\RNum}[1]{\uppercase\expandafter{\romannumeral #1\relax}}

\begin{document}
\preprint{AIP/123-QED}

\title{Spin wave spectral probing of possible microstates in building-block of macroscopically degenerate artificial spin ice}

\author{Nimisha Arora}
\email{narora1.physics@gmail.com}
\affiliation{Department of Physics, Indian Institute of Technology, Delhi, Hauz Khas, New Delhi 110016, India}

\author{Pintu Das}
\affiliation{Department of Physics, Indian Institute of Technology, Delhi, Hauz Khas, New Delhi 110016, India}

\date{\today}

\begin{abstract}
We have investigated the spin wave modes of strongly dipolar coupled, highly anisotropic nanoislands forming square artificial spin ice system using micromagnetic simulation in M{\scriptsize U}M{\scriptsize AX}3 in combination with Matlab coding. Artificial spin ice is considered to be formed by the four square ring-type structure of elliptical cross-section nanoislands. Our results state the direct relation between the spin wave modes generated and the micro-states formed in the system. We have shown that single ring type structure can alone be adequately used to understand the spin wave modes of square artificial spin ice.
\end{abstract}

\pacs{}

\maketitle 
\section{Introduction}
Square artificial spin ice (S-ASI) is the 2-dimensional realization of spin ice behavior observed in 3-dimensional oxide pyrochlore compound such as Dy$_2$Ti$_2$O$_7$, Ho$_2$Ti$_2$O$_7$, etc~\cite{bramwell2001spin}. Such designer nanostructured materials offer possibilities to investigate variety of fascinating phenomena such as physics of frustration~\cite{wang2006artificial}, emergent magnetic monopoles~\cite{morgan2011thermal}, collective magnetization dynamics of interacting spins~\cite{jungfleisch2017high}, and phase transition~\cite{kapaklis2012melting}, etc. Recently, there is a growing interest in the spin  wave behavior in S-ASI due to multiple macroscopically degenerate microstates at the ground state. These microstates offer a route to manipulate spin wave excitation which make these structures a promising candidate for reconfigurable magnonic crystals~\cite{iacocca2016reconfigurable, mamica2012spin}.
\begin{figure}[h!]
\includegraphics[width=0.8\linewidth]{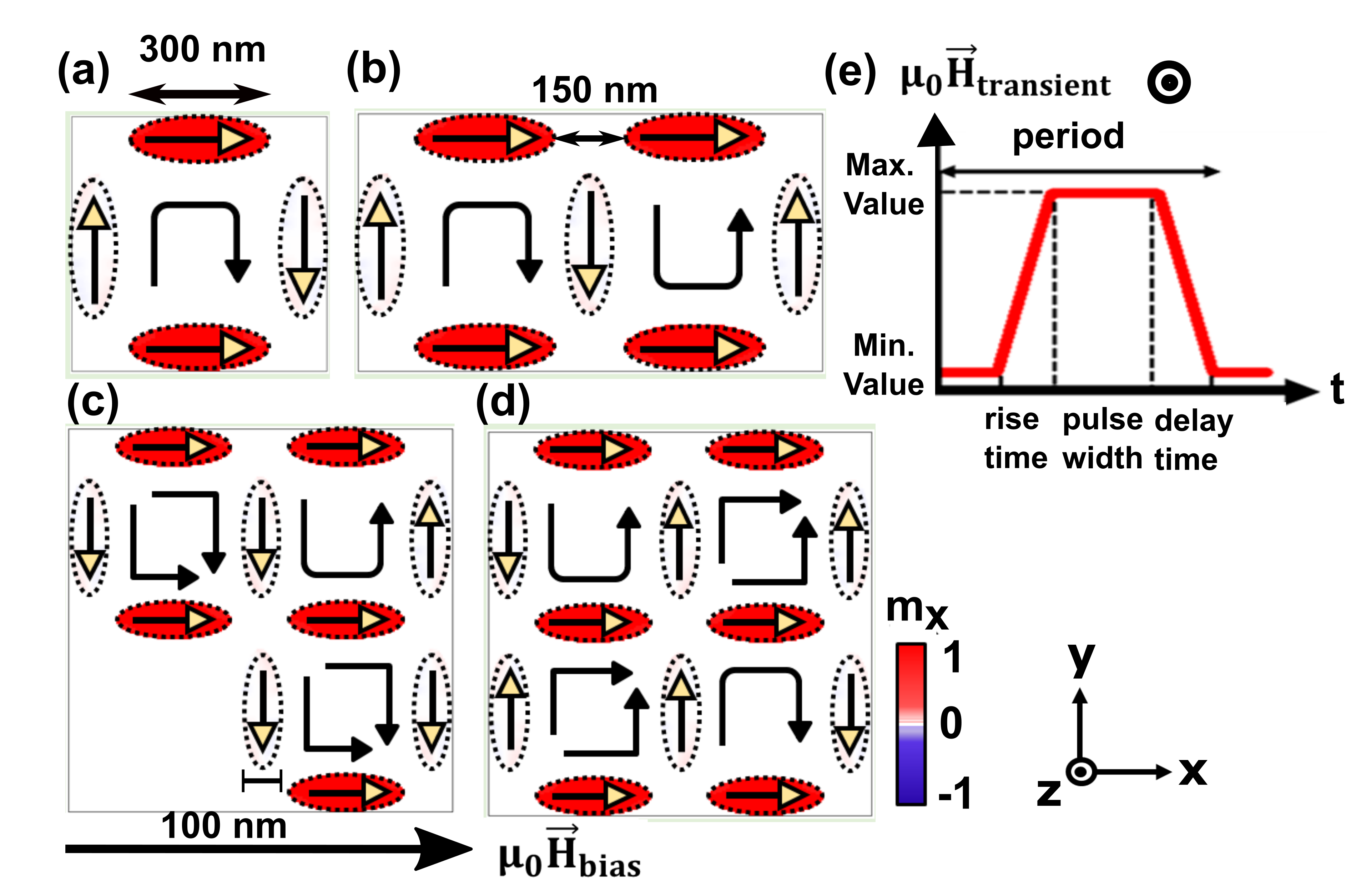}
\caption{remanent micromagnetic configuration for (a) single square ring, (b) double ring (c) triple ring, (d) quadruple ring i.e., S-ASI, and (e) rectangular transient field pulse which is applied along the z-direction to study spin-wave response.}
\label{fig: Intro figure}
\end{figure} 
Moreover, possibilities of locally changing the magnetic behavior of individual nanostructures, such as by the stray field at the magnetic tip of magnetic force microscopes, allows tuning of magnonic band structure of such systems. Indeed, recent studies indicate that ASI can be conceived as analogs of magnonic crystals with tunable band structures offering useful functionalities for applications such as in logic devices~\cite{nikitin2015spin, arava2018computational, arava2019engineering}. Therefore, these structures received a great deal of consideration due to their potential in reprogrammable magnonic crystals~\cite{iacocca2016reconfigurable, mamica2012spin}, computation, data storage~\cite{wang2016rewritable}, Spin logic gate~\cite{nikitin2015spin, arava2018computational, arava2019engineering}, microwave filter~\cite{huynen1999novel}, etc.\\
In particular, experimental and theoretical studies have been performed to understand the magnetization dynamics in S-ASI and it is found that spin wave modes or magnons excited in the systems are affected by the defects~\cite{bhat2016magnetization}, thickness of the nanoislands~\cite{li2017thickness}, topological excitations (emergent magnetic monopoles)~\cite{gliga2013spectral}, etc. Recent studies in this direction shows the possibility of realizing spin wave channels~\cite{iacocca2020tailoring}, manipulation of spin wave response~\cite{arroo2019sculpting, lasnier2020magnetic}, influence of reversal order at the vertices ~\cite{montoncello2018mutual}, etc. 
However, the question of how the underlying magnetic state or the dipolar coupling among the nanostructures affects the spin wave behavior in such ASI systems is still far from being properly understood. In this work, we report the results of investigation on how spin wave modes are effected due to the magnetization reversals of single-domain nanostructures obtained by performing micromagnetic simulations. We also investigated the influence of different magnetic microstates evolved due to reversals on the spin wave properties of the system. Our results show that the overall spin wave behavior of individual vertices of S-ASI with closed edges can be understood from the studies of a square shaped ring of dipolar coupled nanomagnets. Considering such rings as building blocks of individual S-ASI vertices, we find that parallelly placed islands after the magnetization switching shows unique signature of low frequency spin wave modes localized at the switched islands. Hence, our results can be used to probe the evolution of microstates in dipolar coupled ring type structures.
\section{Method}
For the Micromagnetic simulation, we used finite difference discretization based open-source GPU-accelerated software M{\scriptsize U}M{\scriptsize AX}3~\cite{vansteenkiste2014design}. Here, Landau-Lifshitz equation (Eq.\,\ref{eq: LLG}) is employed to calculate the evolution of reduced magnetization $\vec{\mathrm{m}}(\vec{\mathrm{r}}, t)$.
\begin{equation}\label{eq: LLG}
   \frac{\partial\vec{\mathrm{m}}}{\partial t} = \gamma_{\text{LL}}\frac{1}{1+\alpha^2}(\vec{\mathrm{m}}\times\vec{\mathrm{B}}_{\text{eff}} + \alpha(\vec{\mathrm{m}}\times(\vec{\mathrm{m}}\times\vec{\mathrm{B}}_{\text{eff}})))
\end{equation}
with $\gamma_{\text{LL}}$ the gyromagnetic ratio (rad/Ts), $\alpha$ the dimensionless damping parameter, and $\vec{\mathrm{B}}_{\text{eff}}$ the effective field (T).
Here, $\vec{\mathrm{B}}_{\text{eff}} = \vec{\mathrm{B}}_{\text{ext}}+\vec{\mathrm{B}}_{\text{demag}}+\vec{\mathrm{B}}_{\text{exch}}+\vec{\mathrm{B}}_{\text{anis}}+ \vec{\mathrm{B}}(t)+...$\\
where $\vec{\mathrm{B}}_{\text{ext}}$: externally applied field, $\vec{\mathrm{B}}_{\text{demag}}$: magnetostatic field, $\vec{\mathrm{B}}_{\text{exch}}$: Heisenberg exchange field, $\vec{\mathrm{B}}_{\text{anis}}$: anisotropy field, and $\vec{\mathrm{B}}(t)$: time varying magnetic field.\\
For the micromagnetic simulation, structures shown in Fig.\,\ref{fig: Intro figure} are discretized in cuboidal cells of 5\,nm length which is less than the exchange length ($\lambda_{exch} \approx$ 5.3\,nm) of the Permalloy. Experimentally reported value of saturation magnetization $M_{\text{sat}} = 8.6 \times 10^{5}$\,A/m, exchange stiffness constant $ A_{\text{ex}} = 1.3 \times 10^{-11}$\,J/m, and damping coefficient $\alpha = 0.5$ for {Ni}$_{80}$Fe$_{20}$ are used throughout the study~\cite{buschow2003handbook}. For dynamic simulations, the damping coefficient was changed from 0.5 to 0.008, for prolonged precession of weak modes~\cite{barman2009dynamic}.\\
In order to understand the influence of evolved microstates on the spin wave behavior we performed quasi-static dynamics. This dynamics give information about the switching in the nanomagnets which in turn changes the microstates corresponding to the square rings in these specific structures (Fig.\,\ref{fig: Intro figure}). For quasi-static micromagnetic behavior, an external magnetic field is swept within the range of $\pm$\,300\,mT with a field step of 2\,mT applied along x- axis. Magnetization configuration of the geometries is examined at each field value via minimization of total energy using Runge-Kutta (RK45) method. For the study of spin wave dynamics, an additional square time-dependent pulse of magnitude ($\text{B}_{\text{max}}$) of 3\,mT with  rise time ($\text{t}_{\text{rise}}$), duration ($\text{t}_{\text{dur}}$), and fall time ($\text{t}_{\text{fall}}$) of 20\,ps each, was applied along $z$-direction. After the incedence of transient pulse field, magnetization dynamics was observed up to 4\,ns at every 10\,ps time gap. Further analysis is done by performing Fast Fourier transform (FFT) of time evolution of magnetization. Later, corresponding power and phase profiles of the excited spin wave (sw) modes are  mapped by performing FFT of time evolution of magnetic moment at each unit cell of the geometry using MATLAB coding.
\section{Results and Discussions}
\begin{figure}
\includegraphics[width= 0.9\linewidth]{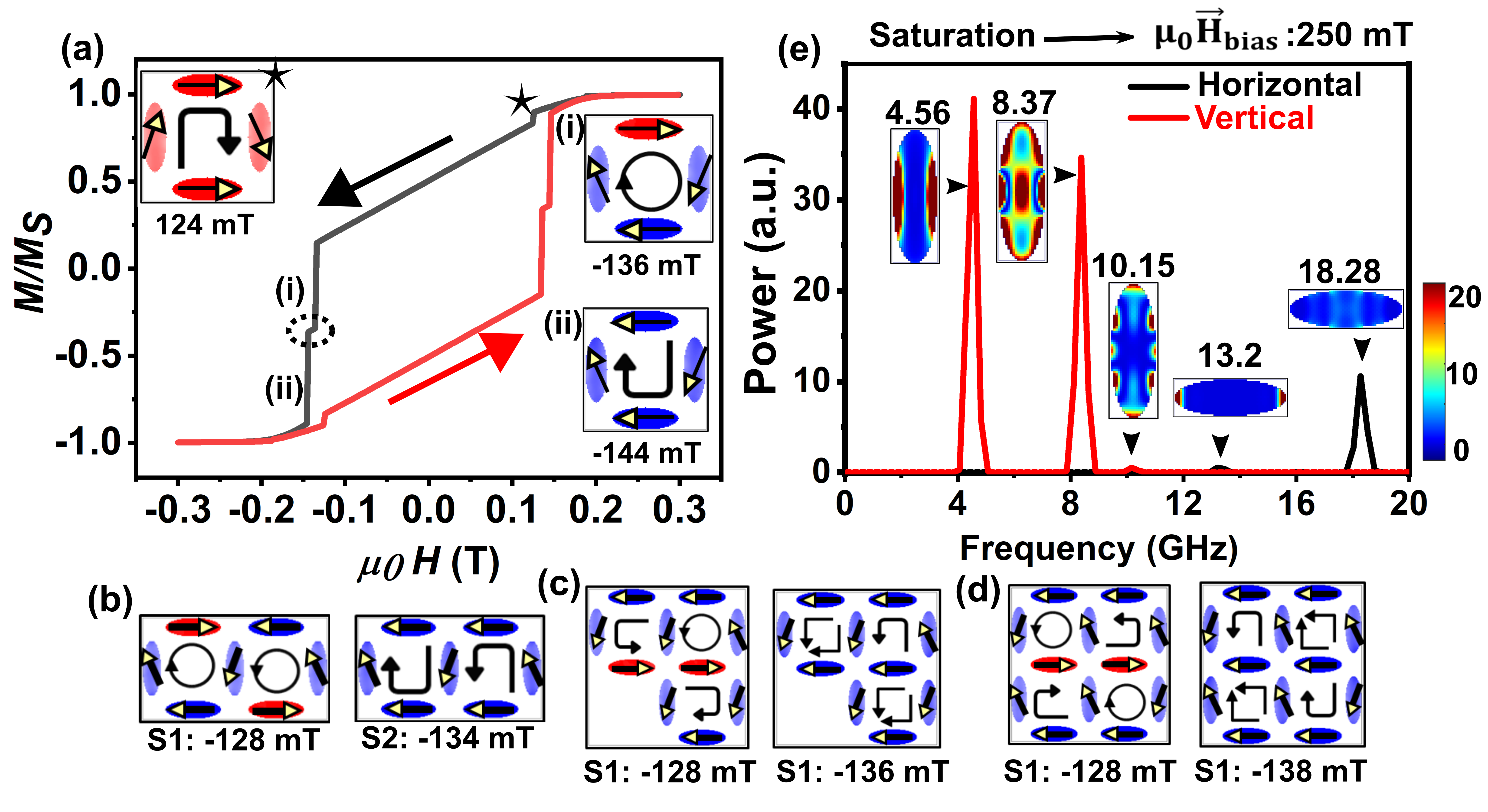}
\caption{Static magnetization dynamics of (a) single ring. Inset shows magnetization configuration in square ring after first (S1) and second (S2) switching. Step observed at 124\,mT is due to significant drop in magnetization ($\sim$30kA/m) due to $\pi/2$ rotation of net magnetization in vertical islands. Evolution of microstates after first (S1) and second  switching (S2) in (b) double ring, (c) triple ring, and (d) quadruple ring e.g., S-ASI. Macrospin directing along the direction of average magnetization is used in magnetization configuration for clarity. (e) spin wave spectra for individual horizontal and vertical island at saturation along x-direction ($\mu_0$H = 250\,mT). Inset shows power profiles of excited mode above their respective peaks.}
\label{fig: hysteresis 1}
\end{figure}

In S-ASI and other ring type structures, arrangement of magnetization configuration in constituting nanoislands decide the microstates present in these systems. Switching the magnetization direction in any constituting nanoisland transform their previous microstates. To investigate the effect of these emerging microstates on their respective spin wave dynamics, input external field parameter have been procured from the quasi-static magnetization dynamics to achieve the desirable microstates. The results shown in Fig.\,\ref{fig: hysteresis 1}(a-d) indicate the cycle of microstate's evolution for structures displayed in Fig.\,\ref{fig: Intro figure}.
For single square ring, nanoisland's magnetization switched by the evolution of microstates from horseshoe to micro-vortex to horseshoe [Fig.\,\ref{fig: hysteresis 1}(a) insets].
These switchings in the nanoislands appeared as steps in the hysteresis curve of the respective structure while the slope of the curve denotes the magnetization behavior in the vertical nanoislands for which easy-axis $\perp \text{B}_{\text{ext}}$. Observed evolution of microstates can be understood in terms of the total energy of the magnetic system which depends on relative orientation of magnetization (\( \mathcal{E}\)$_{\text{head-to-tail}}$ \textless\, \( \mathcal{E}\)$_{\text{head-to-head}}$) direction in neighbouring islands~\cite{keswani2018magnetization}. Since, square ring and other extended ring structures are first saturated along the x-axis, therefore after removing the field all the horizontal islands remain aligned along +ve x direction due to high shape anisotropy and thus exhibit horseshoe microstate at remanent. Similarly, other ring-type structures follow microstate path which ensure maximum head to tail configuration to retain it's minimum energy state. When the number of rings increases beyond 2, onion microstate appeared in the dynamics. Square rings exhibiting onion state at remanent follow $\text{onion}\rightarrow \text{horseshoe} \rightarrow \text{onion}$ microstate's path. Also, results show that both triple and quadruple ring conserve one of the degenerate type \RNum{2} state at the vertex during quasi-static dynamics as reported~\cite{keswani2018magnetization}.\\
\begin{figure}[h!]
\includegraphics[width=0.8\linewidth]{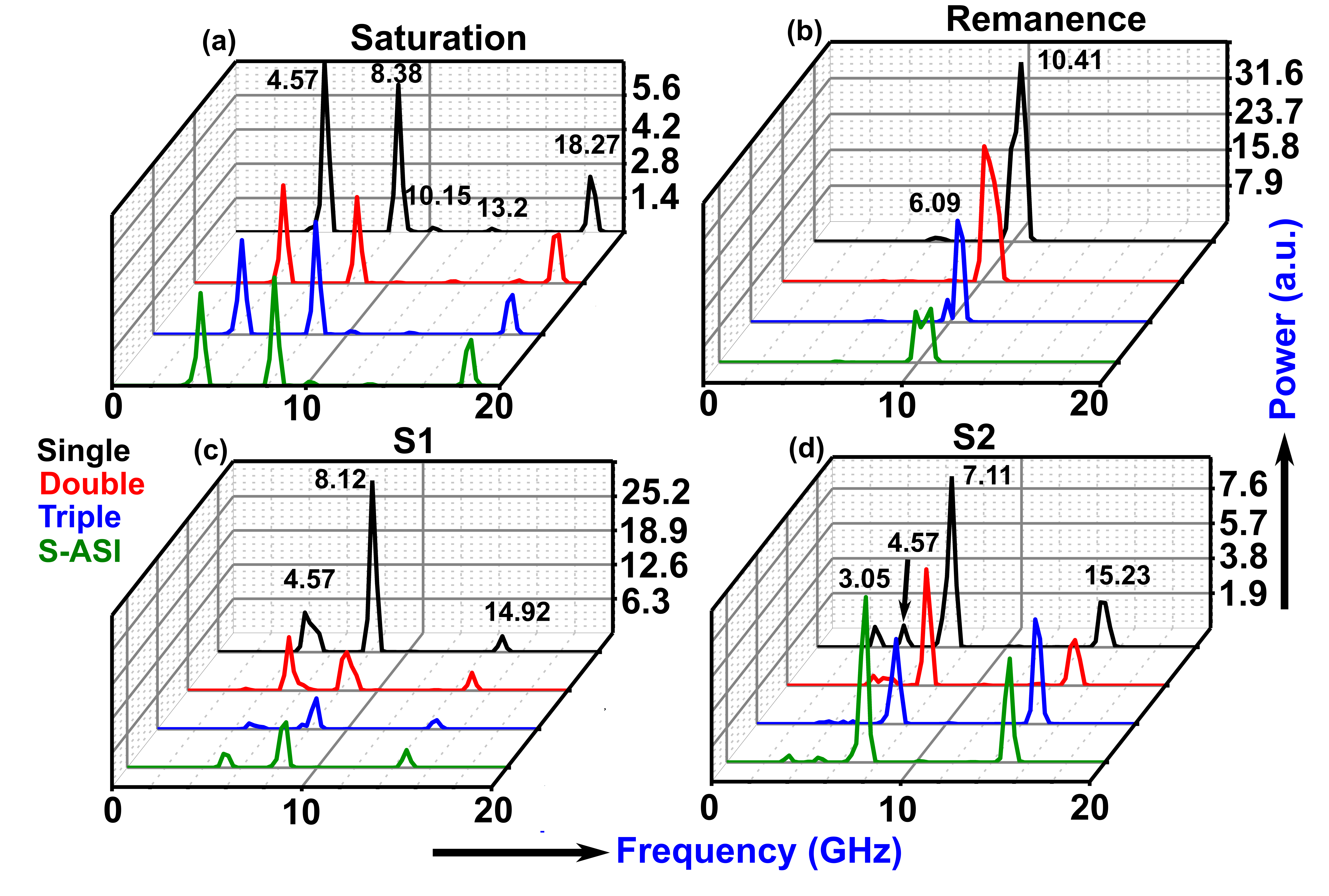}
\caption{Spin wave spectra of the ring type structures shown in (a) saturation, (b) remanence (see Fig.\,\ref{fig: Intro figure} for microstates at remanence), (c) state: S1, and (d) S2 [see Fig.\,\ref{fig: hysteresis 1}(a-d) for microstates at S1 and S2].}
\label{fig: study2_fig1}
\end{figure}
Further to identify the spin wave modes corresponding to individual islands in selective microstates, we have first examined the spin wave modes excited to single nanoisland in saturation. Fig. \,\ref{fig: hysteresis 1}(e) shows the spin wave spectra in horizontal and vertical nanoisland with associated mode power profiles above respective spectral peaks. For the horizontal nanoisland, bulk mode at 18.28 GHz and edge mode at 13.2 GHz were observed as reported in the literature~\cite{gubbiotti2005spin} which indicate the reliability of the simulation performed for other structures. In vertical island, edge mode appear at 4.56 GHz and bulk mode at 8.37 GHz respectively [Fig.\,\ref{fig: hysteresis 1}(e)]. An extra high frequency mode with diminishing amplitude appear at 10.15 GHz which carries the nature of higher order end mode possibly due to the localization of exchange dominated spin wave modes in the region of inhomogeneous internal field. For vertical island, easy axis makes $90^{\circ}$ angle to the applied field which results in the increase of the demagnetization field which consequently reduces the effective field in the nanoisland and thereby decreasing the mode frequency ($\omega_L = \gamma_{LL}\text{B}_{\text{eff}}$).
Thus, we observe low frequency modes for vertical islands and relatively high frequency modes for horizontal island at high bias field [see Fig.\,\ref{fig: hysteresis 1}(e)].\\
\begin{figure}[h!]
\includegraphics[width = \linewidth]{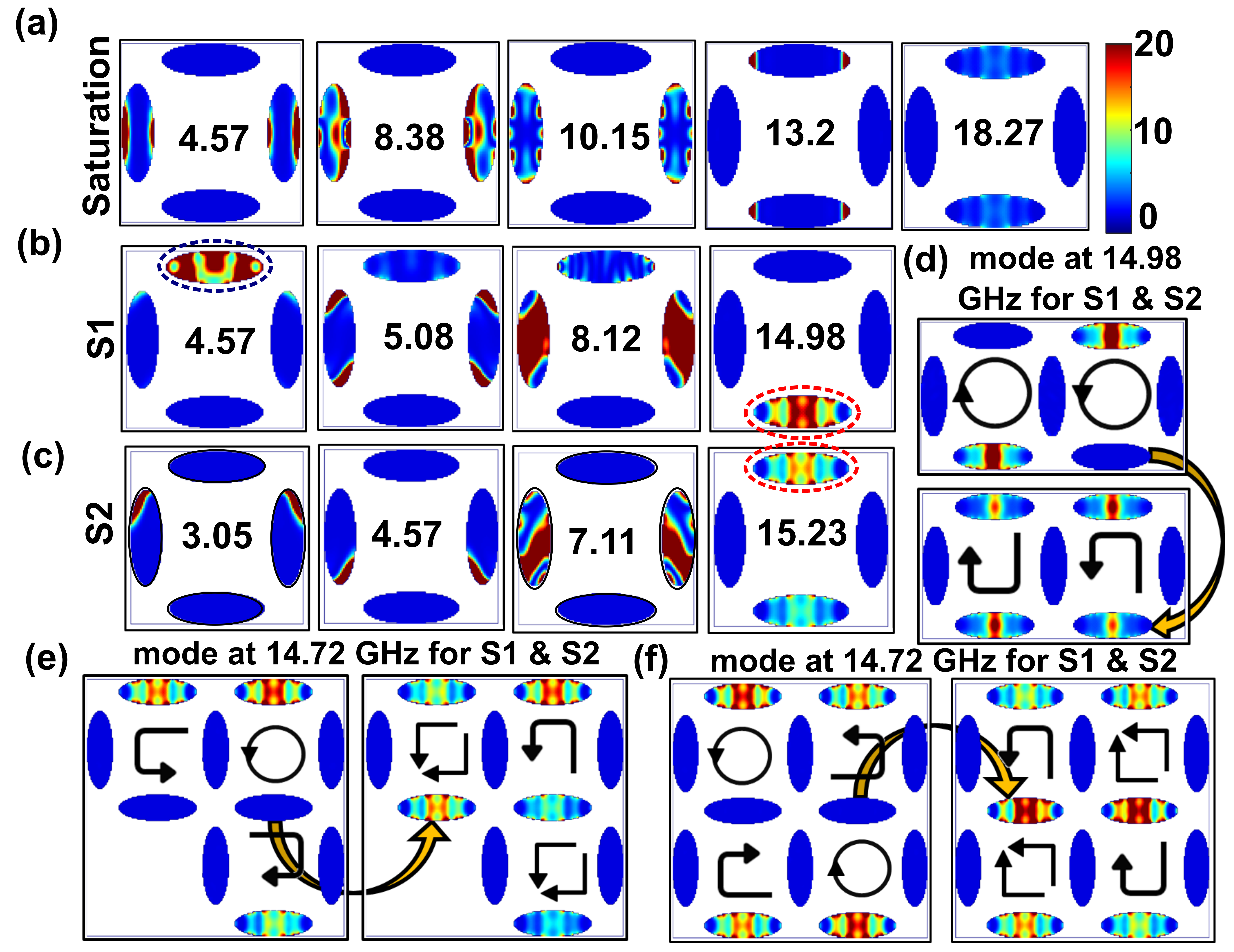}
\caption{Power profile of the excited sw modes in square single ring in (a) saturation, (b) after 1st switching (S1), (c) after second switching (S2). red dotted ellipse encapsulate the switched island whereas blue dotted ellipse encapsulate unswitched island. High frequency sw mode ($\sim$ 14.7\,GHz) at S1 and S2 (identifier mode) for (d) double ring, (e) Triple ring, and (f) Quadruple ring (S-ASI) shows mode localization in switched horizontal islands. Arrows denote switched island from S1 to S2 state.}
\label{fig: study 2 sw of ring 1 & 2}
\end{figure} 
We have then systematically investigated the sw modes of all the ring type structures for four different magnetic configurations (Fig.\,\ref{fig: study2_fig1}) which include saturated state ($+$ve x-axis), remanent state, microstates formed after first and second switching (S1 and S2) as observed in quasi-static magnetization dynamics [Fig.\,\ref{fig: hysteresis 1}(a-d)]. We observed that all the ring type structures including S-ASI in saturation show a strong correlation in spin wave spectra which is evident from the results shown in Fig.\,\ref{fig: study2_fig1}(a). In saturation, five modes found to appear for all structures with three strong modes corresponding to outer-end mode (OE1, 4.57 GHz), bulk mode (B, 8.38 GHz) confined in vertical islands, and bulk mode (B, 18.27 GHz) confined in horizontal islands [see Fig\,\ref{fig: study2_fig1}(a),\,\ref{fig: study 2 sw of ring 1 & 2}(a)]. Corresponding power profiles of excited modes are similar for all the structures therefore power profile of only single square ring is shown in Fig.\,\ref{fig: study 2 sw of ring 1 & 2}(a). In this case, effect of magneto-static interaction is insignificant to affect the magnetic moment distribution in neighbouring island. It can be seen from the sw spectral behavior as it shows the signature of all the five sw-modes localized in individual horizontal and vertical islands at saturation [see Fig.\,\ref{fig: hysteresis 1},\,\ref{fig: study2_fig1}(a)].\\
\begin{figure}[h!]
\includegraphics[width = 0.9\linewidth]{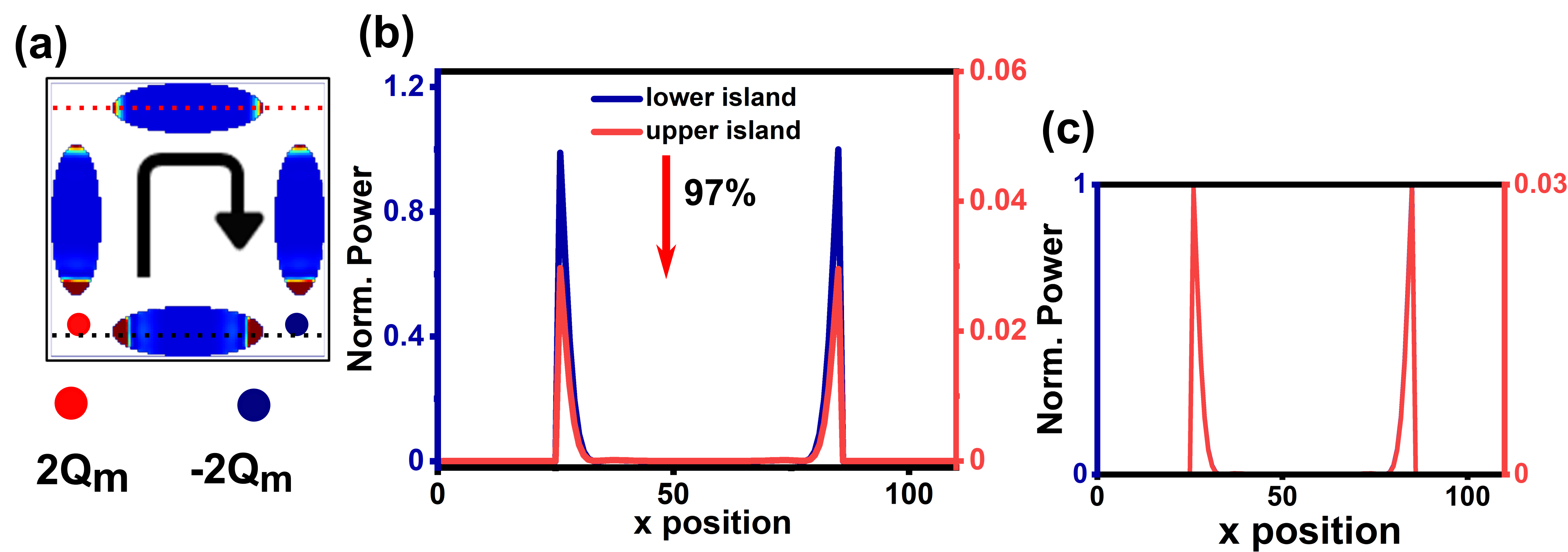}
\caption{(a) Power profile of the OE mode at 6.09\,GHz for square single ring at remanence [see Fig.\,\ref{fig: Intro figure}(a)] (b, c) Power line profile across the center of the upper and lower horizontal island.}
\label{fig: magnetic charge mapping}
\end{figure} 
Now, for the microstates available in remanence, each contributing island has magnetization along it's easy axis. Therefore, both horizontal and vertical islands behave similarly under uniform transient field excitation which results in only two modes [Fig.\,\ref{fig: study2_fig1}(b)]. Here, lower frequency mode ($\approx$ 6.09 GHz) corresponds to the outer end mode and higher frequency mode ($\approx$ 10.41 GHz) corresponds to bulk mode. In remanence, bulk mode appears with a shoulder peak or shows bifurcation in S-ASI. We have analyzed this behavior as a function of island's separation (lattice constant: 300\,nm to 1000\,nm) and our results suggest that either bifurcation or shoulder peak is not present (not shown). Hence, this behavior is due to magnetostatic interaction among the islands. However, power profile corresponding to lower frequency outer end mode shows power variation at the corners which is found to carry the information of the net magnetic charge at the junction. According to dumb-bell model, each single domain nanoisland can be considered as a dipole of magnetic charge $\pm Q_M$ separated by a certain distance (length of the nanoisland). For single square ring, perpendicularly placed island can have head to head (tail to tail, $\pm Q_M$) and head to tail (tail to head, $Q_M\sim0$) spin configuration. Accordingly, lower horizontal island in the single square ring for horseshoe microstate at remanence [Fig.\,\ref{fig: Intro figure}(a)] behave as a dipole of contributing magnetic charge $\approx \pm 2Q_M$. Remaining three nanoislands have closed magnetic flux line distribution with zero magnetic charge at the junction. Close observation of the power-profile shown in Fig.\,\ref{fig: magnetic charge mapping}(a) denotes that power of the outer end modes is mostly concentrated in the corners with nonzero net magnetic charge. Line power profiles at the center of the horizontal islands shown in Fig.\,\ref{fig: magnetic charge mapping}(b) indicates that power associated with OE mode drops to 97\% in the region of zero net magnetic charge as compared to charged region. However, the power distribution profile exactly matches in all the corners [Fig.\,\ref{fig: magnetic charge mapping}(c)].\\
Finally, for the case of S1 and S2 state, sw power spectra [Fig.\,\ref{fig: study2_fig1}(c,d)] shows extensive variation in spectral nature as well as it's corresponding modes profile (Fig.\,\ref{fig: study 2 sw of ring 1 & 2}[b-e]). A weak mode localized in unswitched horizontal island appears at 4.5 GHz. This mode lies near the end mode in vertical islands and therefore shows hybridization. However, we observe OE modes are oriented at certain angle with respect to the easy axis. Our analyses suggest that this orientation is directed along the net magnetization direction due to lower effective field region at the edges. In square single ring, high frequency sw mode ($\sim$ 14.72\,GHz) is localized in lower island which correspond to switched island (observed from micromagnetic simulation, see Fig.\,\ref{fig: hysteresis 1}(a) inset). Thus, high frequency bulk mode of the horizontal islands show marked power variation with respect to unswitched islands [Fig.\,\ref{fig: study 2 sw of ring 1 & 2}(b-f)]. This behavior is consistent in all the ring type structures, therefore, this mode can be considered as switching identifier mode.
The observed variation in sw modes' frequencies for all the states (Fig.\,\ref{fig: study2_fig1}) is due to variation in applied external field to access desirable microstates. Also, power spectra shown in Fig.\,\ref{fig: study2_fig1} clearly indicates that all the sw modes are excited at the approximately same frequencies for all the structures in every case. Corresponding power profiles shown in Fig.\,\ref{fig: study 2 sw of ring 1 & 2} for the excited sw modes are consistent in all the structures. Thus, our results suggest that in order to study the SASI, investigating a single square ring is suffice.\\
In summary, we have studied the spin wave response in four different magnetic configurations for square single ring, double ring, triple ring, and quadruple ring (SASI) structure. We have reported the presence of ubiquitous high frequency switching identifier mode and mapping of the local magnetic charges in terms of power profiles associated to excited outer end modes in remanence. Hence, our results suggest the possibility of employing spin wave modes to probe the switching and evolved microstates in SASI and ring type of structures. 

\begin{acknowledgements}
We acknowledge High Performance Computation (HPC) facility of IIT Delhi. N.A. is thankful to the Council of Scientific Industrial Research (CSIR), Government of India for research fellowship.
\end{acknowledgements}
\section*{DATA AVAILABILITY}
The data that support the findings of this study are available from the
corresponding author upon reasonable request.

\bibliography{aipresearch}

\end{document}